\documentclass[twocolumn,english,aps,prl,reprint,superscriptaddress]{revtex4-1}
\usepackage[T1]{fontenc}
\usepackage[latin9]{inputenc}
\usepackage{xcolor}
\usepackage{amsmath}
\usepackage{graphicx}
\PassOptionsToPackage{normalem}{ulem}
\usepackage{ulem}
\usepackage{subscript}



\usepackage{babel}
\begin{document}

\title{Optically driven collective spin excitations and magnetization dynamics
in the N\'eel-type skyrmion host GaV\textsubscript{4}S\textsubscript{8} }

\author{P. \surname{Padmanabhan}}
\email{ppadmana@ph2.uni-koeln.de}
\address{Physics Institute II, University of Cologne, 50937 Cologne, Germany}

\author{F. \surname{Sekiguchi}}
\address{Physics Institute II, University of Cologne, 50937 Cologne, Germany}

\author{R. B. \surname{Versteeg}}
\address{Physics Institute II, University of Cologne, 50937 Cologne, Germany}

\author{E. \surname{Slivina}}
\address{Physics Institute II, University of Cologne, 50937 Cologne, Germany}

\author{\\V. \surname{Tsurkan}}
\address{Institute of Applied Physics, MD 2028, Chisinau, Republic of Moldova}
\address{Department of Physics, Budapest University of Technology and Economics
and MTA-BME Lendület Magneto-optical Spectroscopy Research Group,
1111 Budapest, Hungary}

\author{S. \surname{Bord\'acs}}
\address{Department of Physics, Budapest University of Technology and Economics
and MTA-BME Lendület Magneto-optical Spectroscopy Research Group,
1111 Budapest, Hungary}
\address{Hungarian Academy of Sciences, Premium Postdoctoral Program, 1051
Budapest, Hungary}

\author{I. \surname{K\'ezsm\'arki}}
\address{Department of Physics, Budapest University of Technology and Economics
and MTA-BME Lendület Magneto-optical Spectroscopy Research Group,
1111 Budapest, Hungary}
\address{Experimental Physics V, Center for Electronic Correlations and Magnetism,
University of Augsburg, 86159 Augsburg, Germany}

\author{P. H. M. \surname{van Loosdrecht}}
\email{pvl@ph2.uni-koeln.de}
\address{Physics Institute II, University of Cologne, 50937 Cologne, Germany}

\begin{abstract}
GaV\textsubscript{4}S\textsubscript{8} is a multiferroic semiconductor
hosting magnetic cycloid (Cyc) and N\'eel-type skyrmion lattice (SkL)
phases with a broad region of thermal and magnetic stability. Here,
we use time-resolved magneto-optical Kerr spectroscopy and micro-magnetic
simulations to demonstrate the coherent generation of collective spin
excitations in the Cyc and SkL phases driven by an optically-induced
modulation of uniaxial anisotropy. Our results shed light on spin-dynamics
in anisotropic materials hosting skyrmions and pave a new pathway
for the optical control of their magnetic order. 
\end{abstract}

\maketitle
The optical manipulation of topologically-nontrivial phases in quantum
materials \cite{Hsieh2008,Yu2011,Adams2012,Liu2014,Lv2015,Xu2015}
is an emerging area within condensed-matter physics \cite{Basov2017},
with efforts aimed at uncovering novel phases and exploring their
non-equilibrium properties. Seminal examples, in this regard, include
the realization of Floquet-Bloch states resulting from photon\textendash surface-state
hybridization in topological insulators \cite{Wang2013,Rechtsman2013}
and helicity-dependent control of topological-surface currents \cite{Kastl2015}.
Interest has also extended to magnetic topological defects known as
skyrmions (Sks) \cite{Nagaosa2013}, fueled by their importance in
memory technology \cite{Fert2013,Tomasello2014,Yu2017}, spintronics
\cite{Rosch2017}, and emergent electromagnetism \cite{Schulz2012,Neubauer2009,Ritz2013}.
Recently, optical stimulus has been successfully used to write and
erase individual Sks \cite{Berruto2018}, confirm their topological
robustness \cite{Langner2017}, and identify new metastable Sk states
\cite{Eggebrecht2017}. 

Skyrmions can be broadly classified into two varieties by their internal
structure. Whirl-like Bloch-type Sks are typically found in chiral
magnets \cite{Berruto2018,Tonomura2012,Adams2012} and are generally
stable over a relatively narrow temperature range in bulk crystals.
N\'eel-type Sks, where spins rotate in radial planes from their cores
to their peripheries, have been identified in bulk lacunar spinels
\cite{Kezsmarki2015,Bordacs2017}, tetragonal oxides \cite{Kurumaji2017},
and thin-film heterostructures \cite{Pollard2017}. Notably, these
systems all posses a polar, rather than chiral, structure and exhibit
axial symmetry. Moreover, N\'eel-type skyrmion lattice (SkL) states
in bulk crystals of these polar magnets show an enhanced thermal
stability. This stems from their orientational confinement primarily
due to the pattern of the Dzyaloshinskii-Moriya interaction (DMI). 

The ultrafast-optical manipulation of magnetic states can generally
be accomplished through mechanisms that leverage direct spin-photon \cite{Zhao2004,Iida2011,Tzschaschel2017}
coupling as well as those that exploit the thermal response of the
host material \cite{Kirilyuk2010,Shibata2018a}. At present,
however, SkL states have only been coherently excited opto-magnetically
using the inverse-Faraday effect in the insulating Bloch-type\textendash SkL-host
Cu\textsubscript{2}OSeO\textsubscript{3} \cite{Ogawa2015}, owing
to its strong linear\textendash magneto-optical response \cite{Versteeg2016}.
Another possible avenue is the optical modulation of magnetic interactions
(e.g. the uniaxial anisotropy), which has proven successful in driving
spin precessions in a variety of magnetic materials \cite{Kimel2004,Bigot2005,Wang2007,Hashimoto2008,Shibata2018}.
Within this context, lacunar spinels, possessing large uniaxial anisotropies
of the easy-axis or easy-plane varieties \cite{Kezsmarki2015,Bordacs2017,Ehlers2016,Ehlers2017},
represent attractive targets for the optical control of SkLs mediated by
the energetic exchange between the electronic, lattice, and spin subsystems. 

In this letter, we report on ultrafast time-resolved magneto-optical
Kerr effect (TR-MOKE) measurements that demonstrate the generation
of coherent collective excitations of the magnetic cycloid (Cyc) and
SkL states in the lacunar spinel GaV\textsubscript{4}S\textsubscript{8}.
Our results reveal GHz oscillations of the magnetization, driven by
a laser-induced thermal modulation of the uniaxial anisotropy. Additionally,
we observe a photo-induced enhancement of the magnetization that originates
from the light-driven switching between the Cyc and ferromagnetic
phases. These experiments establish an alternative route towards the
optical control of the dynamic magnetic character of novel spin textures,
leveraging the intimate coupling between the lattice and spin degrees
of freedom.

\begin{figure}
\includegraphics[scale=0.38]{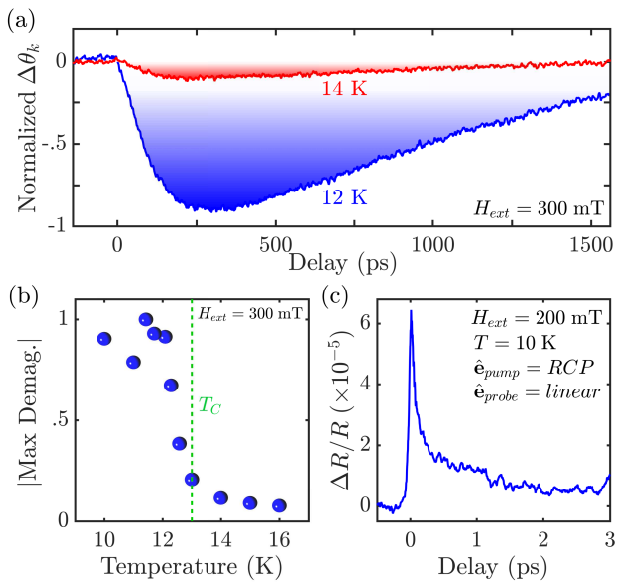}
\caption{\label{fig1}(a) Pump-induced change in the normalized Kerr rotation angle of the
probe pulse below (blue) and above (red) $T_{C}$, (b) the absolute
value of the maximum demagnetization as a function of temperature
and normalized to the value at $T=11.4\text{ K}$, and (c) the differential
reflectivity trace taken at $T=10\text{ K}$.}
\end{figure}

GaV\textsubscript{4}S\textsubscript{8} is a multiferroic narrow-gap
semiconductor belonging to a family of lacunar spinels consisting
of an FCC lattice of tetrahedral (GaS\textsubscript{4})\textsuperscript{5-}
and cubane (V\textsubscript{4}S\textsubscript{4})\textsuperscript{5+}
clusters, the latter carrying $S=1/2$ spins. Below $T_{JT}=44\text{ K}$,
a rhombohedral ($C_{3v}$) distortion appears, with the rhombohedral
axis oriented along one of the cubic body diagonals \cite{Wang2015,Hlinka2016}.
For $T<T_{C}\approx13\text{ K}$ and moderate external fields, the
material is an easy-axis ferromagnet with spins oriented along the
rhombohedral axis. At lower fields, a complex magnetic-phase diagram
emerges consisting of Cyc and SkL ground states due to the competition
between the Heisenberg-exchange interaction, the DMI particular to
the $C_{3v}$ point-group, and the magnetocrystalline anisotropy \cite{Bogdanov1994,Kezsmarki2015}. 

Unlike SkLs in chiral magnets, the N\'eel-type SkL in GaV\textsubscript{4}S\textsubscript{8}
is pinned to the plane perpendicular to the rhombohedral axis \cite{Kezsmarki2015,White2018}.
This is primarily due to the Lifshitz-invariants that comprise the
DMI term, which energetically favor magnetic modulations with $\mathbf{q}$-vectors
perpendicular to the rhombohedral axis. Due to this and the uniaxial
magnetocrystalline anisotropy, the field-stability range of the Cyc
and SkL phases depend on the orientation of the magnetic field with
respect to the rhombohedral axis, since different domains often coexist
in these crystals. As a result, several Cyc and SkL phases can be
supported simultaneously, owing to the different projections of the
external field along the easy axis for the four different domains. In this
work, the external magnetic field was oriented perpendicular to an
as-grown $\left(001\right)$ surface of a GaV\textsubscript{4}S\textsubscript{8}
crystal. This ensured that all the rhombohedral domains were magnetically
equivalent, thereby hosting Cyc and SkL phases over the same field
ranges \cite{Kezsmarki2015}.

We employed TR-MOKE spectroscopy to probe the magnetization dynamics
of the ferromagnetic, Cyc, and SkL states. The pump and probe pulses
($30\text{ fs}$, $1.57\text{ eV}$) were modulated at $100\text{ kHz}$
and $1.87\text{ kHz}$ and focused to $50\text{ \ensuremath{\mu}m}$
and $35\text{ \ensuremath{\mu}m}$ diameters, respectively, with an
on-sample pump fluence of $0.67\text{ \ensuremath{\mu}J/cm}^{2}$.
The magnetic field and sample temperature were controlled with a superconducting-magnetic
cryostat equipped with a variable-temperature insert. To detect the
Kerr-rotation (KR) in the reflected probe beam, we used a polarization-sensitive
bridge, the differential signal from which we measured directly at
the intermodulation frequencies via a phase-sensitive-detection scheme.
This allowed for rapid data acquisition, avoiding the response-time
issues associated with cascaded lock-in-amplifier configurations.

\begin{figure}
\includegraphics[scale=0.38]{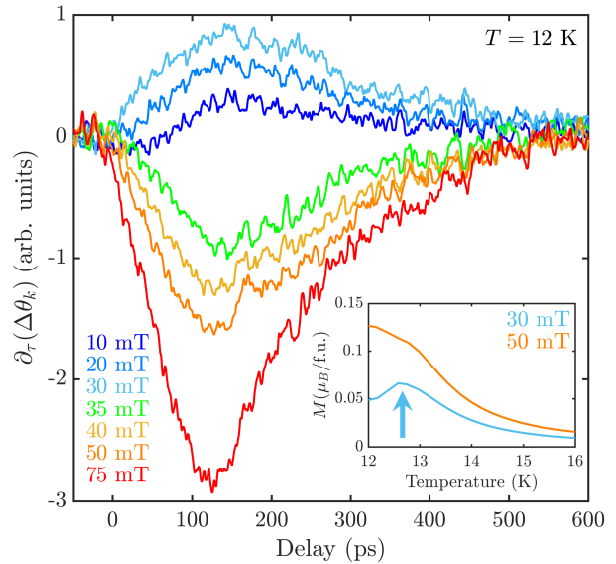}
\caption{\label{fig2}Time-derivative of the TR-MOKE signal at $T=12\text{ K}$ at various
external magnetic fields. The inset shows the temperature dependence
of the magnetization for $H_{ext}=30\text{ mT}$ in cyan and $50\text{ mT}$
in orange. The peak in the former, marked by the arrow, occurs at
the boundary of the Cyc phase.}
\end{figure}

Figure \ref{fig1}(a) shows the pump-induced change to the normalized KR angle
($\Delta\theta_{k}$) of the probe pulse for the ferromagnetic phase
in blue and the paramagnetic phase in red. As seen in Figure \ref{fig1}(b),
the magnitude of the demagnetization step dramatically increases below
$T_{C}$, consistent with the onset of ferromagnetic order. We found
that the demagnetization occurs over $\sim200\text{ ps}$ for all
external fields and sample temperatures that coincide with the ferromagnetic
phase. This timescale is consistent with other semiconducting ferromagnets
and can be attributed to the slow thermalization of the spin system
due to its coupling to the lattice and isolation from the electronic
bath \cite{Kojima2003,Wang2006}. This is supported by the differential-reflectivity
trace shown in Figure \ref{fig1}(c), which contains contributions from electron-electron,
electron-phonon, and phonon-phonon scattering \cite{Othonos1998,Yu2010},
all of which reach quasi-equilibrium within a few picoseconds. The
change in $\Delta\theta_{k}$ occurs on a much longer timescale, demonstrated
by its relatively small variation during the first few picoseconds.
Accordingly, the thermalization of the phonon bath is a nearly instantaneous
event for the spins and the time scale of the demagnetization is primarily
governed by the strength of the magnon-phonon coupling \cite{Kittel1958,Kimel2002}.

The slow response of the spin-system to changes in the lattice is
further exemplified by the magnetization dynamics just below $T_{C}$.
Figure \ref{fig2} shows the time-derivative of $\Delta\theta_{k}$ at $T=12\text{ K}$
for different external magnetic fields ($H_{ext}$). For $H_{ext}\geq35\text{ mT}$,
we observe demagnetization dynamics consistent with the pump-induced
response of the ferromagnetic phase. However, for $H_{ext}\leq30\text{ mT}$,
the signal increases following the pump pulse. This photo-induced
enhancement of the magnetization originates from the temperature dependence
of the magnetization ($M$) across the magnetic phase boundaries \cite{Ruff2015}.
The inset of Figure \ref{fig2} shows $M(T)$ normal to the $\left(001\right)$
surface for $H_{ext}=30\text{ mT}$ and $50\text{ mT}$. As indicated
by the arrow in the inset, the $30\text{-mT}$ $M(T)$ curve is
peaked at the Cyc/ferromagnetic boundary above $12\text{ K}$. Accordingly,
the enhancement of the magnetization results from the comparatively
slow response of the spins to the impulsive heating of the lattice,
pushing the system along the $M(T)$ peak following the pump excitation.
At higher fields, the peak in $M(T)$ vanishes, as shown in the $50\text{-mT}$
curve in the inset of Figure \ref{fig2}, corresponding to the restoration of
conventional demagnetization behavior.

\begin{figure}
\includegraphics[scale=0.38]{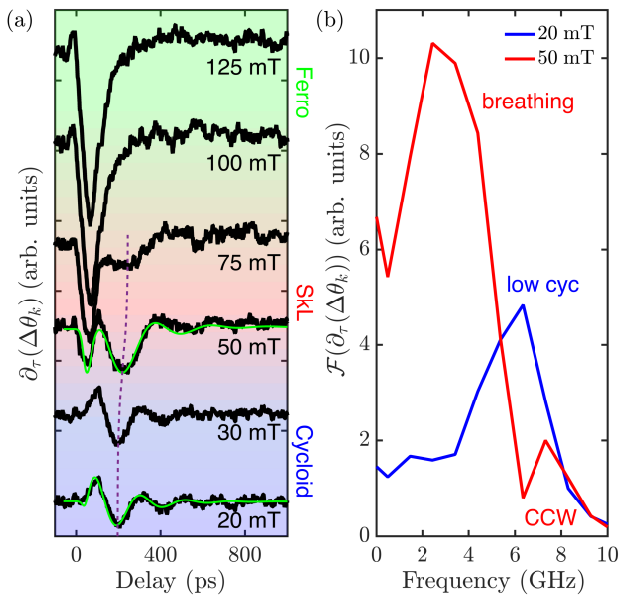}
\caption{\label{fig3}(a) Time derivative of the TR-MOKE signal at $T=10\text{ K}$ for
different external magnetic fields spanning the Cyc, SkL, and ferromagnetic
phases and (b) the Fourier transforms of the $20\text{ mT}$ (blue)
and $50\text{ mT}$ (green) traces in (a). The dashed line in (a)
marks the end of the first oscillatory period. The green lines in
(a) are calculated using Eqs. (\ref{osc}) and (\ref{inco}) with $\beta=0.03$ and $\beta=0.1$
for the $20\text{ mT}$ and $50\text{ mT}$ curves, respectively.}
\end{figure}

To probe the dynamics of the Cyc and SkL phases, we fixed the sample
temperature at $10\text{ K}$ and collected TR-MOKE traces for several
external magnetic fields. The results are shown as black lines in
Figure \ref{fig3}(a), where we plot the derivative of $\Delta\theta_{k}$ to
suppress the incoherent magnetization dynamics. For small fields,
we observe GHz oscillations in the signal, as seen in the $20-75\text{ mT}$
traces. With increasing field, the oscillatory frequency changes,
as noted by the dashed line that marks the end of the first oscillatory
period. For $H_{ext}>75\text{ mT}$, the oscillatory structure vanishes.
Figure \ref{fig3}(b) shows Fourier transforms of two traces in Figure \ref{fig3}(a),
one representative of the Cyc phase ($20\text{ mT}$) and the other
of the SkL phase ($50\text{ mT}$). In the Cyc phase, we see a single
peak centered at approximately $6\text{ GHz}$. Comparing this to
ESR measurements \cite{Ehlers2016}, we identify this as the low frequency
Cyc eigenmode. In the SkL phase, we observe a strong peak centered
at $3.75\text{ GHz}$ and a weaker peak at $7.50\text{ GHz}$.
These frequencies are consistent with the SkL breathing mode and the
counter-clockwise (CCW) rotation mode, respectively \cite{Ehlers2016}. Notably,
the clockwise (CW) rotation mode is absent in our measurements.

We now address two fundamental questions: (1) what is the underlying
mechanism driving the coherent collective spin excitations and (2)
why are only certain modes excited? To answer the first, we note that
the presence of coherent Cyc and SkL excitations were found to be
invariant to the incident pump polarization. This is typically a fingerprint
of a thermal process that does not involve a direct coupling between
the pump-photon field and the spin system \cite{Kirilyuk2010}. Thermal
mechanisms of this type have been explored in the study of laser-induced
spin-precessions in materials such as TmFeO\textsubscript{3} \cite{Kimel2004},
Co films \cite{Bigot2005}, and GaMnAs \cite{Wang2007,Hashimoto2008}. 

Being a polar semiconductor, the electron and optical phonon subsystems
in GaV\textsubscript{4}S\textsubscript{8} are strongly coupled,
leading to a substantial increase in the lattice temperature following
the pump pulse \cite{Othonos1998}. This can, in turn, lead to a modulation
of the magnetocrystalline anisotropy \cite{Kirilyuk2010}. Though
such an effect typically requires a large pump fluence, this constraint
is eased in GaV\textsubscript{4}S\textsubscript{8} due to the strong
temperature variation of its first uniaxial anisotropy constant ($K_{1}$)
below $T_{C}$ \cite{Ehlers2017}. Therefore, the laser-induced heating
of the sample significantly modulates the effective field acting on
the magnetic system through the anisotropy contribution, driving the
magnetic excitations of the SkL and Cyc states. Owing to the relatively
long time required for heat to dissipate from the photo-excited volume
through diffusion \cite{Othonos1998}, this can be interpreted as
a step-like modulation of $K_{1}$ within the experimental window.

To justify the above description, we used the finite-difference time-domain
method to solve the Landau-Lifshitz-Gilbert equation for the SkL state
using the \texttt{Mumax3} code \cite{Vansteenkiste2014}. The effective
field acting on the magnetic system is given by
\begin{eqnarray}
\mathbf{H}_{eff} & = & \mathbf{H}_{ext}+\mathbf{H}_{ani}+\mathbf{H}_{d}+\mathbf{H}_{DMI}+\mathbf{H}_{exch}\nonumber \\
 & = & H_{ext}\hat{\mathbf{e}}_{ext}+\frac{2K_{1}}{\mu_{0}M_{S}}\left(\hat{\mathbf{e}}_{u}\cdot\text{\textbf{m}}\right)\hat{\mathbf{e}}_{u}+\mathbf{H}_{d}\nonumber \\
 &  & \;+\frac{2D}{\mu_{0}M_{S}}\left[\mathcal{L}_{xz}^{\left(x\right)}+\mathcal{L}_{yz}^{\left(y\right)}\right]+\frac{2A_{ex}}{\mu_{0}M_{S}}\nabla^{2}\mathbf{m}\label{Heff}
\end{eqnarray}
where $A_{ex}=0.0588\text{ pJ/m}$ is the exchange stiffness, $M_{S}=28.8\text{ kA/m}$
is the saturation magnetization, $K_{1}=10\text{ kJ/m}^{3}$ is
the (steady state) anisotropy constant, $D=0.043\text{ mJ/m}^{2}$
is the DMI constant, $\mathcal{L}_{jk}^{\left(i\right)}=m_{j}\partial_{i}m_{k}-m_{k}\partial_{i}m_{j}$
are Lifshitz-invariants corresponding to $C_{3v}$ symmetry, $\hat{\mathbf{e}}_{u}$
is a unit vector in the direction of the easy axis, $\hat{\mathbf{e}}_{ext}$
is a unit vector in the direction of the applied field, $\mathbf{m}=\mathbf{M}/M_{S}$,
and $\mathbf{H}_{d}$ is the demagnetizing field. The material parameters
were estimated from literature and match the experimental periodicity
of the SkL state \cite{Kezsmarki2015,Ehlers2017}. Here, $\hat{\mathbf{x}}'\parallel\left[100\right]$,
$\hat{\mathbf{y}}'\parallel\left[010\right]$, $\hat{\mathbf{z}}'\parallel\left[001\right]$,
$\hat{\mathbf{x}}\parallel\left[11\overline{2}\right]$, $\hat{\mathbf{y}}\parallel\left[1\overline{1}0\right]$,
and $\hat{\mathbf{z}}=\hat{\mathbf{e}}_{u}\parallel\left[111\right]$.
We introduced a time dependence in the effective magnetic field through
a step-like decrease of $K_{1}$ by 1\% of its steady state value,
consistent with our estimate of the pump-induced heating of the lattice.
The simulated system consisted of a $128\times64$ grid of $0.8\text{ nm}^{3}$
cuboids with periodic boundary conditions along $\hat{\mathbf{x}}$
and $\hat{\mathbf{y}}$, initialized with one unit cell of a triangular
N\'eel-type SkL with the SkL-plane normal to $\hat{\mathbf{z}}$. The
stability of this state was established by slowly field cooling the
system in the presence of an external magnetic field and fluctuating
thermal field \cite{Vansteenkiste2014,Zhang2017}. The lattice parameters
were then determined by minimizing the total energy. For the results
discussed below, the external field parallel to $\hat{\mathbf{z}}$
was $125\text{ mT}$ and the field along $\hat{\mathbf{x}}$ was
varied, resulting in a tilting of the external field with respect
to $\hat{\mathbf{e}}_{u}$ by an angle $\alpha$.

\begin{figure}
\includegraphics[scale=0.38]{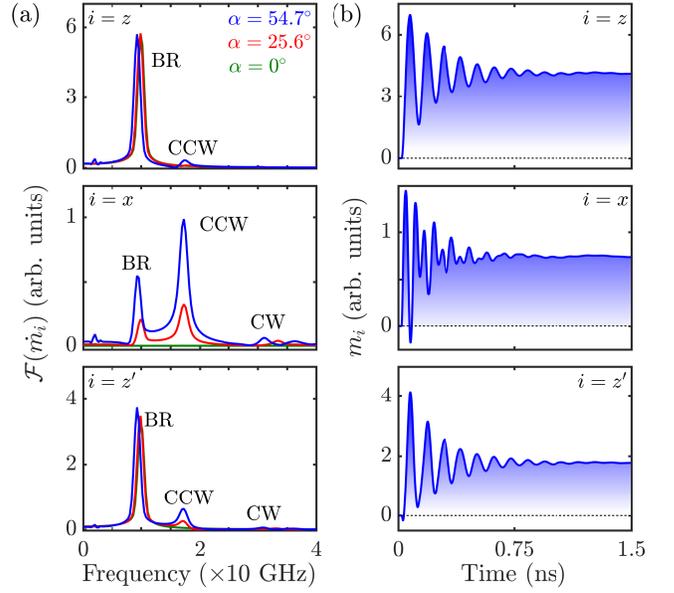}
\caption{\label{fig4}(a) The Fourier transforms of the simulated $\dot{m}_{z}$, $\dot{m}_{x}$,
and $\dot{m}_{z'}$ for various values of $\alpha$ spanned by the
external magnetic field and the rhombohedral axis and (b) the simulated
$m_{z}(t)$, $m_{x}(t)$, and $m_{z'}(t)$ for $\alpha=54.7{}^{\circ}$.}
\end{figure}

Figure \ref{fig4}(a) shows the Fourier transforms of $\dot{m}_{z}$, $\dot{m}_{x}$,
and $\dot{m}_{z'}$ for various values of $\alpha$. For $\alpha=0^{\circ}$,
we see a single resonance peak that manifests only in the $z$-component,
corresponding to the breathing mode. This is because the modulation
of $\mathbf{H}_{eff}$ is entirely along the $z$-direction (i.e.
normal to the SkL plane) and can therefore only couple to the breathing
mode \cite{Mochizuki2012}. As $\alpha$ is increased, however, we
see the gradual appearance of two additional peaks appearing in the
$z$-, $x$-, and $z'$-components of the magnetization. The appearance
of the new resonances is due to the the core shift characteristic
of N\'eel-type SkLs subject to oblique external fields \cite{Leonov2017}.
This results in a deformation of the Sk texture, reducing the six-fold
rotational symmetry of the SkL to a two-fold rotation, thereby introducing
a time-dependent component to the effective field in the plane of
the SkL, which activates the rotational modes \cite{Mochizuki2012}.
The tilting of the net magnetization and deformation of the SkL are
relatively small, which accounts for the weakness of the CCW mode
in our experimental results where $\alpha=54.7^{\circ}$. Further,
we see that the third simulated resonance peak is relatively weak,
a fact that is supported by the absence of the CW mode in our measurements.
Finally, we note that the simulated resonances were blue-shifted with
respect to the experimental results. This deviation is most likely
due to the strong sensitivity of the mode frequencies to the values
of $D$ and $A_{ex}$ \cite{Zhang2017}. Nevertheless, the order of
the simulated resonances is consistent with previous ESR measurements
\cite{Ehlers2016} as well as our experimental observations.

We now construct a phenomenological model of the experimental TR-MOKE
traces. From Figure \ref{fig4}(b), we see that the magnetization dynamics resulting
from a modulation of $K_{1}$ are comprised of decaying sinusoidal
oscillations superimposed on a step-like offset. This reflects the
transient reorientation of $\mathbf{m}$ due to the reduced anisotropy
following the optical pump. This type of response can be modeled by
a damped harmonic oscillator driven by a step-like force, in this
case, representing the optically-induced modulation of the uniaxial
anisotropy. The incoherent de/remagnetization dynamics can be described
by the sum of two exponentials convolved with a step-like function
representing the response time of the spin-system to the lattice.
For both the oscillatory and incoherent parts, we use the same step-like
function. Finally, we model the magnetization dynamics as the sum
of the incoherent and oscillatory contributions, taking $\Delta\theta\left(t\right)=A\cdot I_{i}\left(t\right)+B\cdot I_{o}\left(t\right)$
where the incoherent ($I_{i}$) and oscillatory ($I_{o}$) parts of
the signal are given by the solutions to
\begin{equation}
\frac{d^{2}I_{o}}{dt^{2}}+2\gamma\omega_{0}\frac{dI_{o}}{dt}+\omega_{o}^{2}I_{o}=g\left(t\right)\cdot e^{-t/\tau}\:\text{,}\label{osc}
\end{equation}
\begin{equation}
I_{i}\left(t\right)=\left(h\ast g\right)\left(t\right)\label{inco}
\end{equation}
where $h\left(t\right)=e^{-t/\tau_{1}}-\beta e^{-t/\tau_{2}}$ and
$g\left(t\right)=\left[\text{erf}\left(\alpha t\right)+1\right]$.
Here, $\tau_{1}$ and $\tau_{2}$ are the demagnetization and remagnetization
time-constants, $\beta$ is a scaling parameter, $\alpha$ controls
the spin-response time, and $\tau$ is the rate at which $K_{1}$
returns to the pre-time-zero value. Estimating $\tau_{1}=130\text{ ps}$,
$\tau_{2}=\tau=2600\text{ ps}$, and $\alpha=0.05$ from the experimental
results, we obtain the curves plotted in green in Figure \ref{fig3}(a). The
agreement between this model and the experimental results illustrates
that the measured magnetization dynamics reflect the competition between
incoherent and oscillatory signals.

In summary, we have demonstrated the ultrafast optical generation of
coherent collective spin excitations of the Cyc and SkL phases in
GaV\textsubscript{4}S\textsubscript{8}, driven by an optically-induced
modulation of the uniaxial magnetocrystalline anisotropy. This indirect
coupling between the optical pulse and the spin system is mediated
by the lattice and represents a new mechanism by which magnetic excitations
can be generated in skyrmion-host compounds with strong anisotropy.
Furthermore, the peculiar nature of the magnetic ordering at the phase
boundaries of GaV\textsubscript{4}S\textsubscript{8} allows for
a transient enhancement of the magnetization driven by the optically-induced
heating of the lattice. This study underscores the intimate coupling
between the spin and lattice subsystems in GaV\textsubscript{4}S\textsubscript{8},
and may provide a framework for the optical control of topological
magnetic objects in semiconductors.
\begin{acknowledgments}
P.P., F.S., R.B.V., E.S., and P.H.M.vL. acknowledge financial support
from the the Deutsche Forschungsgemeinschaft (DFG) through SFB-1238
(Project B05). V.T. and I.K. acknowledge financial support from the DFG via
the Transregional Research Collaboration TRR 80: From Electronic Correlations
to Functionality (Augsburg-Munich-Stuttgart) and via the Skyrmionics
Priority Program SPP2137. S.B. acknowledges financial support from
the National Research, Development and Innovation Office \textendash{}
NKFIH, ANN 122879, and the BME-Nanotechnology and Materials Science
FIKP grant of EMMI (BME FIKP-NAT).
\end{acknowledgments}

\bibliographystyle{apsrev4-1}
\bibliography{GVSTRMOKE_Manuscript_Final_arXiv_2.bbl}

\end{document}